# Discrete energy levels of Caroli-de Gennes-Martricon states in quantum limit due to small Fermi energy in FeTe$_{0.55}$Se$_{0.45}$

Mingyang Chen[*], Xiaoyu Chen[*], Huan Yang[†], Zengyi Du, Xiyu Zhu, Enyu Wang, and Hai-Hu Wen[†]

Center for Superconducting Physics and Materials, National Laboratory of Solid State Microstructures and Department of Physics, Collaborative Innovation Center for Advanced Microstructures, Nanjing University, Nanjing 210093, China

[*]These authors contributed equally to this work.

Caroli-de Gennes-Martricon (CdGM) states were predicted in 1964 as low energy excitations within vortex cores of type-II superconductors[1]. In the quantum limit, namely $T/T_\mathrm{c} \ll \Delta/E_\mathrm{F}$, the energy levels of these states were predicted to be discrete with the basic levels at $E_\mu$ = ± $\mu\Delta^2/E_F$ ($\mu$ = 1/2, 3/2, 5/2, …). However, due to the small ratio of $\Delta/E_F$ in most type-II superconductors, it is very difficult to observe the discrete CdGM states, but rather a symmetric peak appears at zero-bias at the vortex center[2,3]. Here we report the clear observation of these discrete energy levels of CdGM states in FeTe$_{0.55}$Se$_{0.45}$. The rather stable energies of these states versus space clearly validates our conclusion. Analysis based on the energies of these CdGM states indicates that the Fermi energy in the present system is very

**small.**

**Introduction**

When magnetic field is applied to a type-II superconductor, vortex with quantized flux of $\Phi_0 = h/2e$ = 2.07 × 10$^{-15}$ Wb will be formed. At the center of a single vortex, the superconducting order parameter is zero and it recovers a unified value in the scale of the coherence length $\xi$. Due to the confinement to the quasiparticles by the vortex core, Caroli-de Gennes-Matricon (CdGM)[1] predicted that there are confined low energy bound states with the energy levels at about $E_\mu = \pm \mu\Delta^2/E_F$ ($\mu$ = 1/2, 3/2, 5/2, …) with $\Delta$ the superconducting energy gap and $E_F$ the Fermi energy. However, due to the very small value of $\Delta/E_F$, these discrete energy levels of CdGM states have never been really observed. In most conventional superconductors, the easily observed feature is that a peak of density of states (DOS) locates at zero energy with a symmetric shape[2,3]. This peak will split and fan out when moving away from the center. Later on, it was understood that[4-6] this symmetrized peak around zero energy at the vortex center is due to the accumulated DOS arising from many symmetric CdGM states when the quantum limit situation $T/T_c \ll \Delta/E_F$ is not satisfied. The theoretical calculations starting from the Bogoliubov-de Gennes equations can explain not only the symmetrically shaped bound state peak at Fermi level at the core center, but also the splitting into two peaks which exhibit the fanning out behavior at positions away from the center. Although the bound state peak of the lowest energy level has been argued for the quantum limit in YNi$_2$B$_2$C (ref.

7), YBa$_2$C$_3$O$_{7-\delta}$ (YBCO) (ref. 8), Bi$_2$Sr$_2$CaCu$_2$O$_{8+\delta}$ (Bi-2212) (ref. 9) and iron based superconductors[10-12], while the clear evidence of the discrete CdGM states is still lacking. For example, the vortex core states exhibit as a dominant asymmetric peak near Fermi level in YNi$_2$B$_2$C (ref. 7) and iron based superconductors[10-12]. In cuprate superconductors, some kind of symmetric peaks on scanning tunneling spectrum (STS) were observed within the vortex core, but it was argued that these may result from the competing orders[8,9,13]. In this study, we present the clear evidence of the discrete CdGM states in the quantum limit in FeTe$_{0.55}$Se$_{0.45}$.

**Results**

Figure 1b shows the temperature dependence of the mass magnetization of the FeSe$_{0.55}$Te$_{0.45}$ single crystal, and the superconducting transition occurs at $T_c$ = 13.3 K. Figure 1a shows a typical topographic image measured by scanning tunnelling microscopy (STM). One can observe clear atomically resolved topography with the lattice constant of 3.79 ± 0.02 Å in two perpendicular lattice directions, which is consistent with 3.8 Å of the other measurements[14,15]. According to previous investigations[14-17], the brighter spots on the surface represent the Te atoms, while the darker spots are the Se atoms with smaller atomic size. No interstitial Fe atoms with signatures of large bright spots[16] have been observed on the cleaved top surface in our measurements, which manifests that most of the interstitial Fe impurities have been removed by the annealing treatment (see Methods). Figure 1c shows a series of scanning tunnelling spectroscopy (STS) measured along the arrowed line in Fig. 1a at $T$ = 0.4 K. Obviously our measured spectra show a full gap feature everywhere with one or two pairs of superconducting coherence peaks, which is consistent with previous reports[14,16]. The statistics on the peak positions on the same sample is shown in Supplementary Fig. 1. On a specific tunnelling spectrum we can see one or two pairs of coherence peaks with energies varying from 1.1 meV to 2.1 meV. The two gap feature observed on one tunneling spectrum indicates a multiband feature, and the particular shape of the spectrum strongly depends on the local details of the structure. When a magnetic field of 5 T is applied, vortices can be observed on the sample by measuring the zero-bias conductance map as shown in Fig. 1d. One can see that the vortex lattice is disordered and is obviously very different from the ordered vortex lattice observed on $Ba_{0.6}K_{0.4}Fe_2As_2$ (ref. 10). The disordered vortices are also observed in $FeTe_{1-x}Se_x$ in previous study (ref. 12). These vortices are pinned by some regions of weak superconductivity or local defects, and it is consistent with the inhomogeneous electronic properties reflected by the spatial inhomogeneity of the tunnelling spectra shown in Fig. 1c. If we enlarge the view on an individual vortex, we can find that many of them have anisotropic or irregular shapes in real space. This seems to be a common feature in both cuprate and iron based superconductors. Although we did not observe the clear evidence of checkerboard electronic state within the vortex core as observed in $Bi_2Sr_2CaCu_2O_{8+\delta}$ (ref. 18) and $Ca_{2-x}Na_xCuO_2Cl_2$ (ref. 19), the shape of an individual vortex is strongly affected by the inhomogeneity of surface electronic structure, this manifests that the superconductivity may be unconventional.

**Discussion**

In order to study the vortex-core state, we measured the tunnelling spectra across a single vortex. In Fig. 2a, we show the d$I$/d$V$ mapping measured at zero bias. One can clearly see a single vortex with a roughly round shape. Following the parallel traces marked by the three white arrowed lines in Fig. 2a, we measured the tunnelling spectra across the vortex with increment steps of 3.8 Å, and show the results in Fig. 2b-d, respectively. One can see that the tunneling spectra measured far away from the vortex center are similar to those measured at 0 T. Near the center of the vortex, the spectra reveal two remarkable features. Firstly, the superconducting coherence peaks are suppressed completely, indicating a suppression of order parameter in this region. Secondly, there are some isolated bound-state peaks, and those close to the zero-bias exhibit roughly a symmetric shape. The peaks near the zero-bias locate actually at +0.5 ± 0.05 meV and -0.6 ± 0.05 meV, respectively. The slight difference of the peak energy at positive and negative bias may be induced by the local impurities or disorders[20]. The vortex-core states in FeTe$_{0.55}$Se$_{0.45}$ observed here are apparently different from those of conventional superconductors, e.g., 2$H$-NbSe$_2$ (ref. [2]) in which a giant peak was observed symmetrically around zero-bias and it fans out when moving away from the vortex center. The feature of discrete bound state peaks vanishes near the edge of vortex core. We argue that these peaks are the vortex bound states in the quantum limit. When we focus on the bound state peaks at positive-bias side, there are three discretized peaks between the zero-bias and the larger superconducting gap. If we attribute the vortex core states to CdGM type in the quantum limit, the three peaks locating at +0.45, +1.2, and +1.8 meV may refer to the low-lying bound states with $\mu$ = 1/2, 3/2, and 5/2, respectively. Considering the first level at positive bias $E_{1/2} = \Delta^2/2E_F \approx 0.45$ meV, combining with the superconducting gap $\Delta$ from 1.1 to 2.1 meV, we can obtain the Fermi energy $E_F$ which ranges from 1.3 to 4.9 meV. The calculated value of Fermi energy is really very small and comparable to the superconducting gap. The small Fermi energies determined here are very close to those determined from angle-resolved photoemission spectroscopy (ARPES) measurements[21,22]. The measurements along three vertical traces are also done for the same vortex; the results are similar and shown in Supplementary Fig. 2.

To further elaborate the properties of discrete energy levels in the vortex core, we extract the energy values and intensities of the three discrete peaks at positive bias and show them in Fig. 3a and b as a function of distance along the trace marked by the dark arrowed line in the inset of Fig. 3a. The original spectra with offset for increment spatial step of 3.8 Å are shown in Fig.3c. It is clear that the peak energies of $E_\mu$ ($\mu$ = 1/2, 3/2, and 5/2) are almost unchanged with the STM tip position, but the intensity of $E_{1/2}$ peaks decreases when the tip moves away from the vortex center. The feature is better shown in Fig. 3d by a two-dimensional colour plot of the spatial evolution of the tunnelling spectra. It is obvious that the peaks of the discrete energy levels near the vortex center are almost independent of positions yielding three parallel ridge-like traces in the positive bias side. Such tendency remains until the STM tip moves out of the vortex core region with $\xi \approx 25$ Å. The characteristics of these bound states are consistent with the CdGM states in quantum limit by theoretical predictions[23]. If we plot the spatially dependent amplitude of d$I$/d$V$ taken at bias of $E_{1/2}$ = 0.45 mV for this vortex along six different traces in a wider region, it is easy to see a dip at around +3.82 nm and -3.44 nm. Outside the dip, a pair of small second peaks of d$I$/d$V$ amplitude appears at about $\pm$ 4.5 nm away from the vortex center for most line cuts on this vortex. The related results are shown in Supplementary Fig. 3a. We argue that this 'dip' and 'second peak' may be the theoretically predicted amplitude oscillation of CdGM states in quantum limit[23], which thus provides another proof and is further discussed in Supplementary Note 2.

Focusing on the spectrum near the vortex core center, as shown by the red line in Fig. 3c, one can see that the spectrum is strongly particle-hole asymmetric, i.e., the differential conductance intensity of the lowest bound state peak on negative energy is larger than that on positive energy. This asymmetric feature seems to be different from the theoretical results for single band system[23] and the experimental results in YNi$_2$B$_2$C (ref. [7]), in both cases the $E_{1/2}$ peak is much stronger than $E_{-1/2}$ peak. We believe that this asymmetric spectrum and different features on the positive and negative energy side are induced by the existence of shallow electron and hole bands with different Fermi energies in the system. One may argue that the second and third peaks on the spectrum at the vortex core center have similar energies of the superconducting gaps far away from the vortex core, thus they could be the "residual" gap feature of the system. This is unlikely and not possible. According to the basic understanding of a vortex structure based on the Ginzburg-Landau theory, all pairing order parameter at the vortex core center should vanish, this leaves no space for a "residual" gap feature with the unchanged gap values.

In addition to the discrete energy levels mentioned above, we also observed some vortices with different patterns of vortex bound states. In some cases, we find the tunnelling spectra showing one dominant but extremely asymmetric bound-state peak locating near zero-bias. Figure 4 shows three sets of tunnelling spectra crossing different vortices observed on different samples with this kind of feature. This type of tunnelling spectrum is consistent with previous observation within the vortex core of iron based superconductors[10-12]. Although the images of three vortices shown in the insets of Fig. 4a-c have slightly different shapes, the main features of these vortex bound states seem to be similar and are very different from those shown in Figs. 2 and 3. From the spatially resolved tunnelling spectra in Fig. 4a-c, one can see that the peak positions of the vortex bound state locate at some positive bias-voltage from 0.2 to 0.6 meV. However, all the peaks do not split or change their peak positions when the tip moves away from the vortex center, which is similar to the situation of discrete bound states. The extremely asymmetric feature of the vortex bound states in these vortices may be induced by the local details of electronic structure. Theoretically, it was predicted that the peak with $\mu = 1/2$ may have very high intensity in the quantum limit in a single band system[23]. The asymmetry of the DOS in the occupied and unoccupied states in iron-based superconductors may also lend an explanation for the extremely asymmetric bound-state peak intensity at positive and negative bias in our present system[15,24]. Since vortices favor to be pinned by the defects or disorders, which may give strong influence on the detailed shape of bound state peaks in the present sample. Statistically, we have conducted measurements on 7 vortices, three of them show the discrete and roughly symmetric energy levels, four show the asymmetric single peak near zero-bias, but none of them shows the peak position exactly at zero. In Supplementary Fig. S4 we present the spatial evolution of tunnelling spectra crossing other three vortices beside those shown in Fig. 2 and Fig. 4. The vortex core states are strongly influenced by the local defect or the surrounding vortices[25], or the asymmetric density of state to Fermi level[26]. These may give explanations for different patterns of bound state peaks in different vortices.

On all the spectra that we measured on seven vortices, we have not observed any bound-state peak which appears exactly at zero-bias. We emphasize that the offset-bias in the STS measurements may affect the exact peak positions, thus a well calibration of the bias-voltage is very important. The offset bias in our experiment is determined from the averaged *I-V* curves, i.e., we use the voltage value for zero tunnelling current as the offset voltage, which means that the current should be zero when the effective bias-voltage is zero. The offset voltage values for zero-tunnelling-currents are +0.15, +0.13, and +0.10 meV for the spectra shown in Fig. 4a, b, and c respectively. This offset effect in our experiment has been well checked and removed. For statistics, we collect the energies of bound state peaks close to the zero bias from all measured seven vortices. The results are shown in

Supplementary Fig. 5. All the lowest bound state peaks locate at energies between 0.2 to 0.8 meV for both positive and negative bias.

For both types of the vortex bound state peaks we witnessed here, the peak energies almost do not shift when the STM tip moves away from vortex center. This is consistent with the theory for the CdGM bound states in the quantum limit, therefore we attribute them all to the CdGM states. In three vortices we observed the discrete CdGM states with roughly symmetric shape of the lowest level bound state peak. In conventional superconductors with large $E_F$, the giant bound-state peak measured at the core center will fan out and move to larger bias-voltage when the tip moves away from the vortex center. However, the vortex bound sate peak energy in quantum limit will not shift by moving of the STM tip on the superconductors with very small $E_F$ compared with $\Delta$. By suing the Pippard relation $\xi_0 = \hbar v_F/\pi\Delta$, it is easy to derive that $\Delta/E_F \sim 1/k_F\xi_0$. As mentioned before, in the quantum limit $T/T_c \ll \Delta/E_F$. In FeTe$_{1-x}$Se$_x$ samples, $k_F$ (or $E_F$) is very small as revealed by ARPES for both hole and electron pockets[21,22]. We did the measurements at about 400 mK, and $T/T_c \approx$ 0.03 which is much smaller than $1/k_F\xi_0 \approx$ 0.3 to 0.6 [using $\xi_0$ = 25 Å, $k_F$(α-band) = 0.07 Å$^{-1}$, $k_F$(β-band) = 0.12 Å$^{-1}$ (ref [21,22,27])], with $\xi_0$ the coherence length in zero-temperature limit. Therefore, this analysis gives strong argument for the existence of discrete CdGM states in quantum limit in present system. As we mentioned already, none of our tunnelling spectra exhibits the bound state peak exactly at zero, this is different from a recent observation which claims the possible evidence of Majorana mode[28]. We do not want to exclude that possibility since only partial bands are involved in forming the possible Dirac cone structure[29] in the material, and perhaps that experiment just successfully detects the surface states of that band. More experiments with well taking of the offset problem are necessary to clarify this issue. Our present experiments provide clear evidence for the discrete CdGM bound states which has been long sought for decades.

**Methods**

**Sample synthesis**.

The FeTe$_{1-x}$Se$_x$ single crystals with nominal composition of $x$ = 0.45 were grown by self-flux method[30]. The excess Fe atoms were eliminated by annealing the sample at 400 °C for 20 hours in O$_2$ atmosphere followed by quenching in liquid nitrogen.

**STM/STS measurements**

The STM/STS measurements were carried out in a scanning tunnelling microscope (USM-1300, Unisoku Co., Ltd.) with ultra-high vacuum, low temperature and high magnetic field. The samples were cleaved in an ultra-high vacuum with a base pressure about 1×10$^{-10}$ torr. Pt/Ir tips were used for all the STM/STS measurements. A typical lock-in technique was used for the tunnelling spectrum measurements with an AC modulation of 0.3 mV and 973.8 Hz. All the tunnelling spectra were recorded in the tunnelling condition of $V_{bias}$ = 10 mV and $I_{set}$ = 500 pA. The offset bias voltages in STS measurements is determined from the averaged *I-V* curves, i.e., we use the voltage value for zero tunnelling current as the offset voltage to make sure that the current should be zero when the bias-voltage is zero. The vortex images are measured by zero-bias conductance mapping.

**Data availability.**

The data that support the plots within this paper and other findings of this study are available from the corresponding author upon reasonable request

**Acknowledgements**

We acknowledge useful discussions with Tao Xiang, Jenny Hoffman, Tetsuo Hanaguri, Christoph Berthod, and Dimitri Roditchev. The work was supported by National Key R&D Program of China (grant number: 2016YFA0300401), National Natural Science Foundation of China (NSFC) with the projects: 11534005, 11374144, and Natural Science Foundation of Jiangsu (grant number: BK20140015).

**Author contributions**

The low-temperature STS measurements and analysis were performed by M.Y.C, X.Y.C, H.Y. and H.H.W. The samples were grown by Z.Y.D and X.Y.Z. H.Y., M.Y.C. and H.H.W contributed to the writing of the paper. H.Y. and H.H.W. are responsible for the final text. All authors have discussed the results and the interpretations.

**Competing financial interests**

The authors declare that they have no competing financial interests.

*Correspondence and requests for materials should be addressed to

huanyang@nju.edu.cn, hhwen@nju.edu.cn

**Figures and captions**

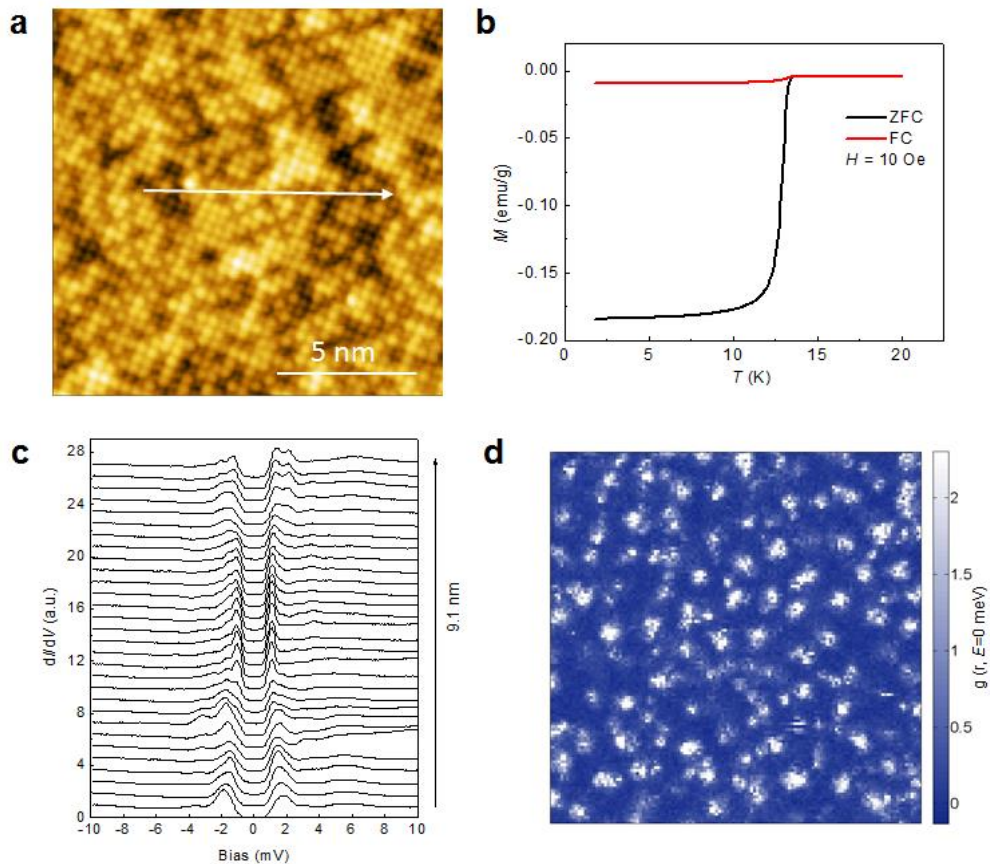

**Figure 1 | STM results and superconducting transition of FeTe$_{0.55}$Se$_{0.45}$. a,** Atomically-resolved topographic image with the square lattice measured at 1.8 K with a bias voltage of $V_{bias}$ = 10 mV and tunnelling current of $I_{set}$ = 300 pA. **b,** Temperature dependence of mass magnetization after zero-field-cooled (ZFC) and field-cooled (FC) modes at 10 Oe. **c,** Spatially resolved tunnelling spectra along the arrowed line measured at 0.4 K. The superconducting gaps seem very inhomogeneous on the sample. **d,** Spectroscopic image of vortex lattice measured by zero-bias conductance map at $T$ = 0.48 K and $B$ = 5 T, the field of view dimensions are 200 nm × 200 nm.

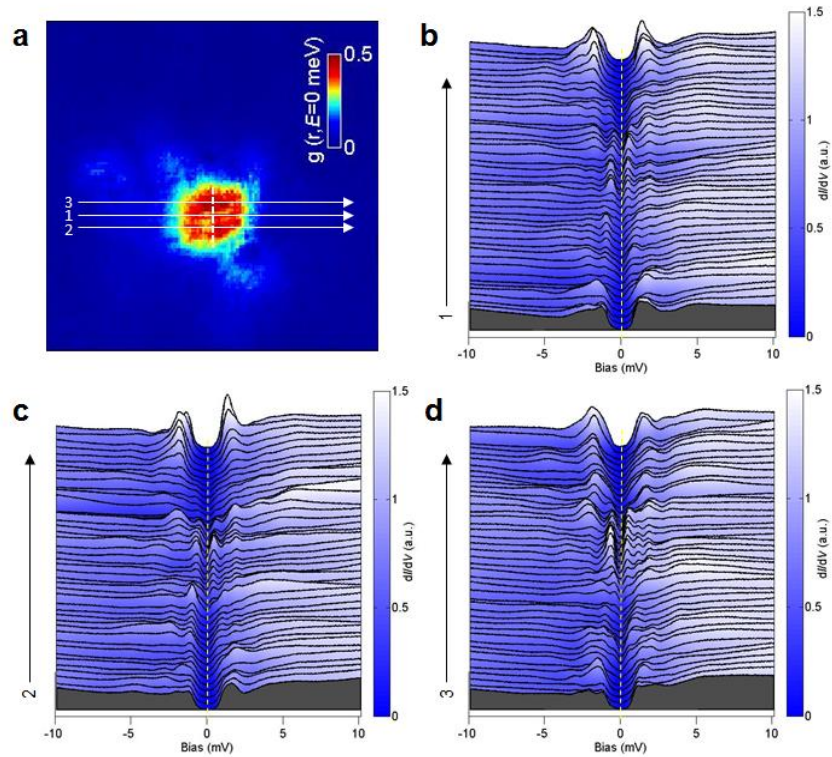

**Figure 2 | Vortex image and CdGM states near the vortex core center. a**, Image of a single vortex in a 20 nm × 20 nm region measured at 0.48 K and 4 T. **b-d**, Tunnelling spectra measured along the arrowed lines marked from 1 to 3 in **a**. The dashed lines in **b-d** show the positions of zero bias voltage for each figure. The CdGM states can be clearly observed near the vortex core center.

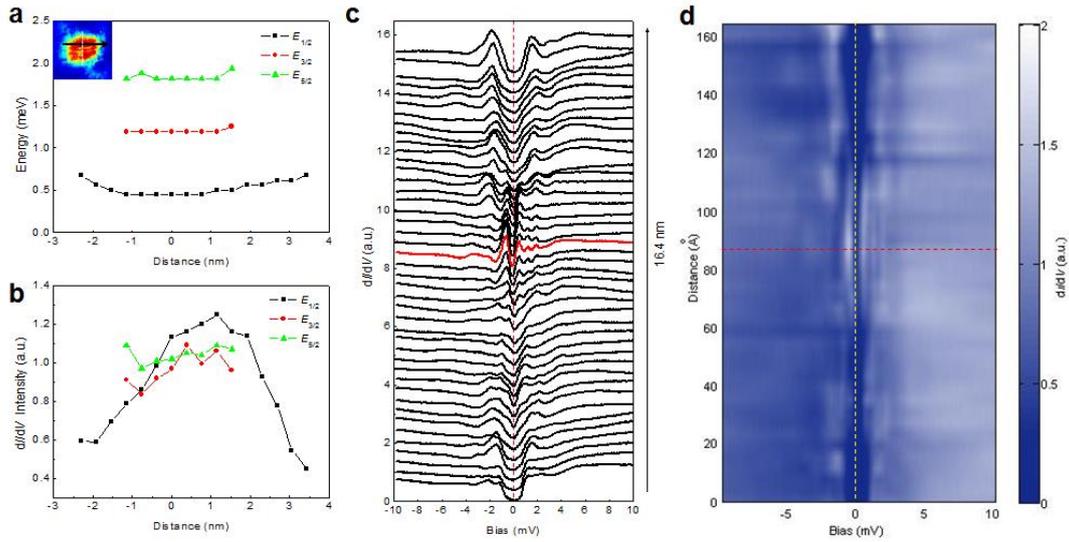

**Figure 3 | Vortex bound state peak positions and intensities. a**, Peak energies of $E_{1/2}$, $E_{3/2}$, and $E_{5/2}$ as function of tip position along the trace marked by the black line crossing a vortex shown in the inset. **b**, The differential conductance intensity of bound state peaks at the peak energies of $E_{1/2}$, $E_{3/2}$, and $E_{5/2}$ as the values in **a**. **c**, Spatially resolved tunnelling spectra measured across the vortex, and the red line represents the spectrum measured at the center of the vortex core. Each curve is offset for an increment distance of 3.8 Å. **d**, Colour plot of spatial profile of the spectra crossing the vortex. The red dashed line represents the position of the vortex core center.

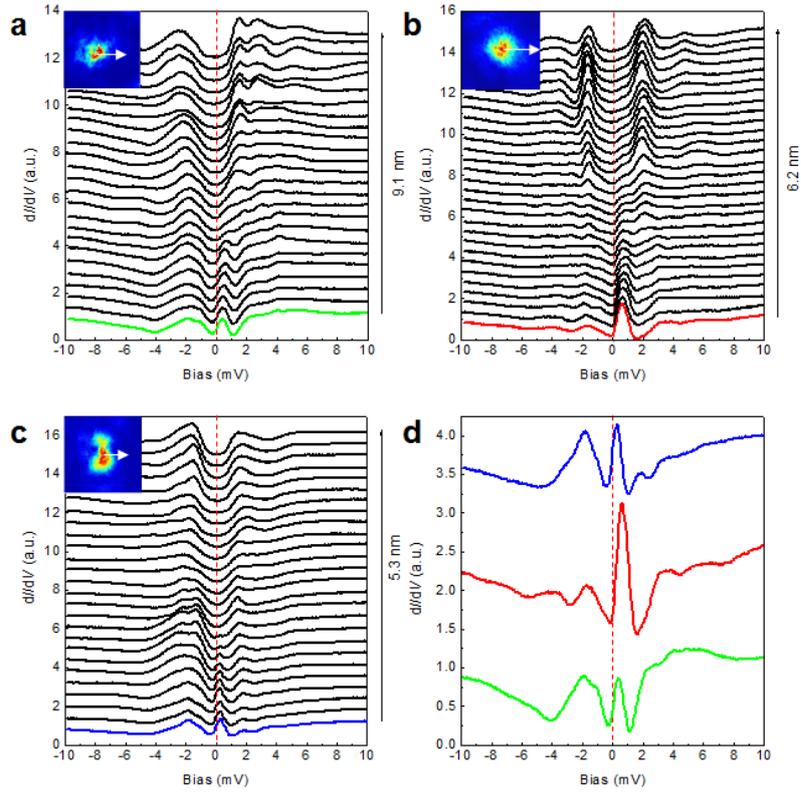

**Figure 4| Vortex core with single bound state peak. a-c**, Series of spatially resolved tunnelling spectra measured at 0.48 K and 4 T along the traces marked by the white lines drawn in the insets respectively. The red, green or blue line in **a-c** represents the spectrum measured at different vortex centers. **d**, Spectra measured at the center of three different vortices. The red dashed lines in all figures show the positions of zero-bias voltage. For these three vortices, there is only one bound state peak appearing at a positive bias near the vortex center.

# Supplementary Information

# Discrete energy levels of Caroli-de Gennes-Martricon states in quantum limit due to small Fermi energy in FeTe$_{0.55}$Se$_{0.45}$

Mingyang Chen[*], Xiaoyu Chen[*], Huan Yang[†], Zengyi Du, Xiyu Zhu, Hai Lin, and Hai-Hu Wen[†]

## 1. Statistics on superconducting gaps

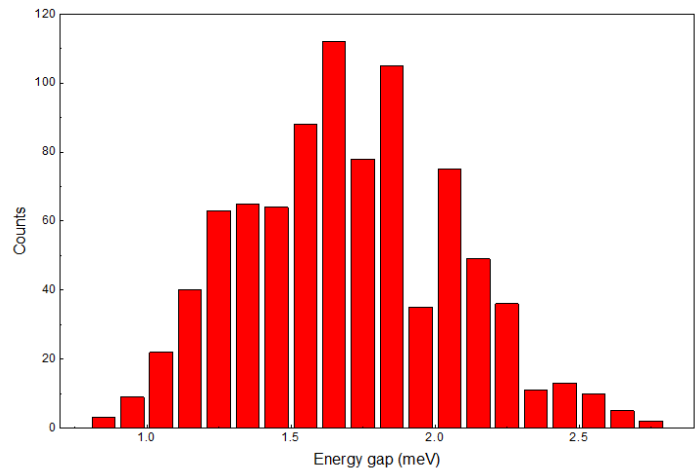

**Figure S1 | Histogram of the superconducting gap $\Delta$ at 0T.** The absolute values of superconducting gaps are determined by the coherence-peak positions for all the 355 spectra measured on one sample. The typical spectra are shown in Fig. 1c. There are usually two pairs of coherence peaks locating at about ±1.1 mV and ±2.1 mV, or one pair of coherence peak locating at about ±1.6 mV.

## 2. Further information of CdGM states in the vortex shown in Fig. 2

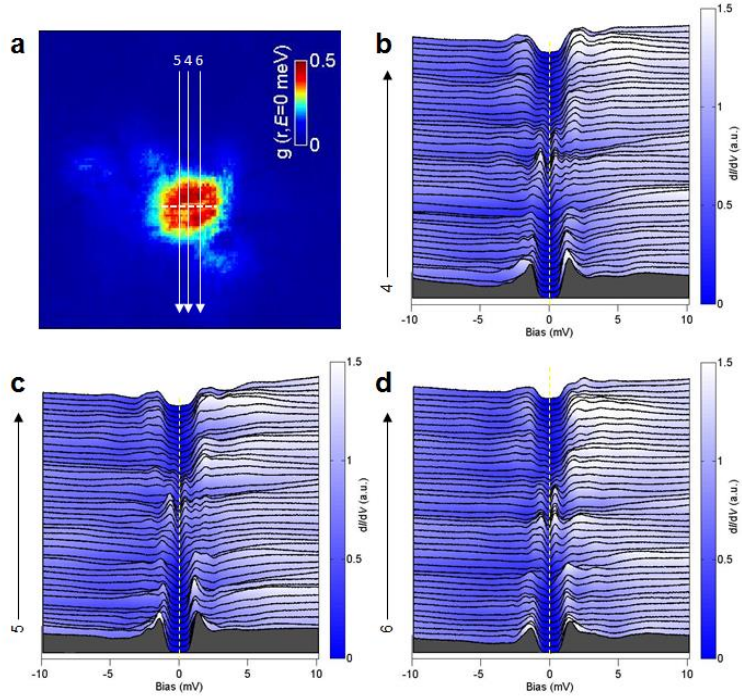

**Figure S2 | Vortex image and CdGM states along three vertical lines. a**, Image of the same vortex as in Fig. 2 in the main text. **b-d**, Tunnelling spectra measured along vertical white lines marked from 4 to 6 in **a**. The dashed lines in **b-d** show the positions of zero bias voltage for each figure.

Figure S3a shows the spatial evolution of the differential conductance amplitude measured at a fixed energy of $E_{1/2}$ = 0.45 mV. One can find that a pair of small second peaks appears at about ± 4.5 nm. Figure S3b shows some spectra measured at some typical positions. One can find clear kinks near 0.45 mV on the spectrum measured at -4.2 nm away from the vortex core center. One possible reason for these small d$I$/d$V$ peaks come from the amplitude oscillations as a function of distance away from the vortex center. Theoretically, the peak amplitude of DOS for one selected bound state will show spatial oscillation in quantum limit[S1,S2]. The space period of the amplitude oscillation is usually related to $1/k_F$. In conventional superconductors, the Fermi energy $E_F$ or the Fermi vector $k_F$ is much larger than $\Delta$ or $1/\xi_0$. Hence, the second-order bound state peak of $E_{1/2}$ exists at somewhere $r_{2nd} < \xi_0$, e.g., $r_{2nd} \approx 0.5\xi_0$ = $4/k_F$ for a material with $k_F\xi_0$ = 8 from the theoretical calculation[S1]. However, $k_F$ is very small in FeTe$_{0.55}$Se$_{0.45}$, i.e., $k_F\xi_0 \approx 1.7$ to 3. In such condition, it is not strange that the second order of $E_{1/2}$ bound state peak exists at the place $r_{2nd} > \xi_0 \approx 25$ Å. We argue that the small d$I$/d$V$ peaks locating at about ± 4.5 nm is from the amplitude oscillation of CdGM state in quantum limit, which is another proof for this state.

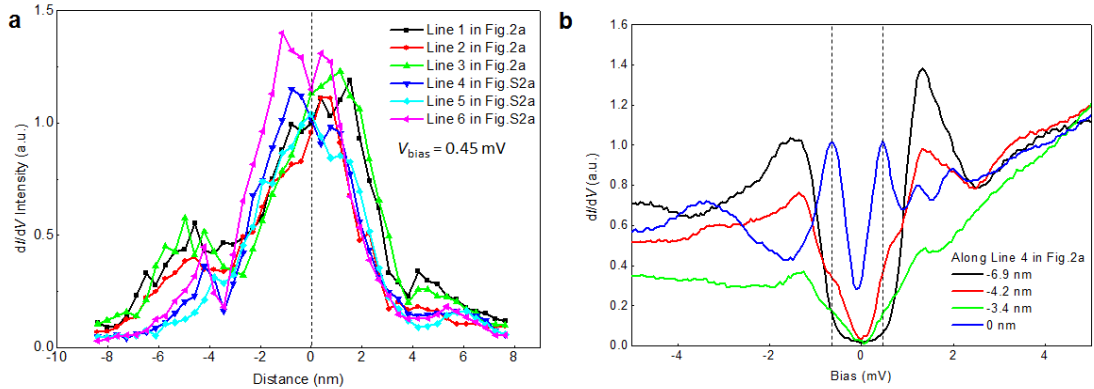

**Figure S3 | Possible second-order vortex bound state peak of $E_{1/2}$. a**, Spatially evolution of the d$I$/d$V$ intensity taken at a fixed bias voltage of $E_{1/2}$ = 0.45 mV. One can find possible second-order peaks locating at about ±4.5 nm. **b**, Tunnelling spectra measured at some typical positions. The dashed lines show $E_{\pm1/2}$ peaks, and there are both kinds with strong d$I$/d$V$ intensity at the same energy on the spectrum measured at -4.2 nm away from the vortex center.

## 3. CdGM states in other measured vortices

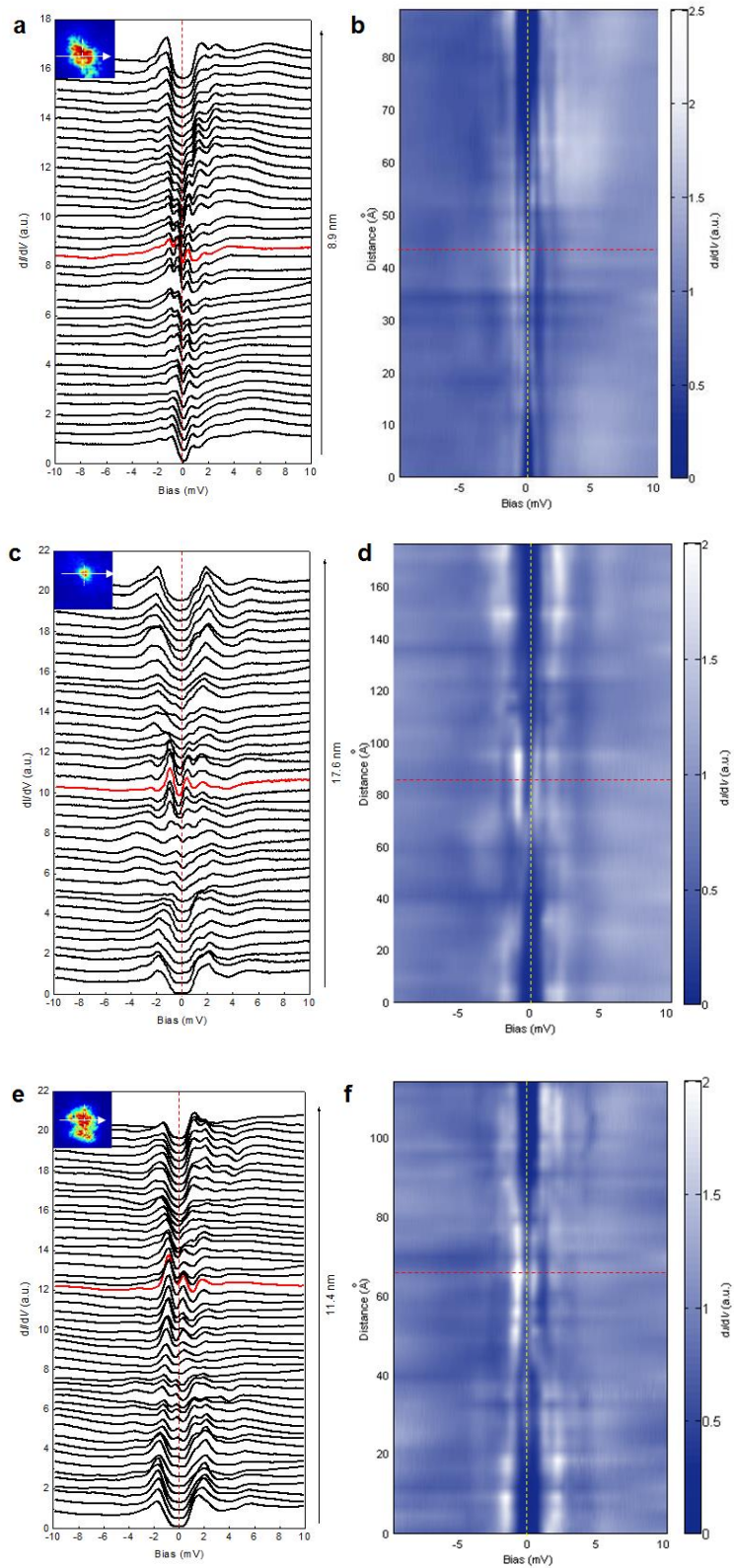

**Figure S4 | Other typical vortex core states in different vortices. a, c, e** Series of spatially resolved tunnelling spectra measured at 0.48 K and 4 T along the traces marked by the white lines drawn in the insets. The red lines in **a**, **c**, and **e** represent

the spectra measured at vortex core center. **b**, **e**, and **f** are the colour plots of spatial profile of the spectra shown in **a**, **c**, and **e**, respectively. The vortex bound states in **a**-**d** are two asymmetric peaks to zero-bias, while the bound state in **e** or **f** is a single peak with peak position at some positive energies..

## 4. Energy distribution of $E_{\pm 1/2}$ peak positions

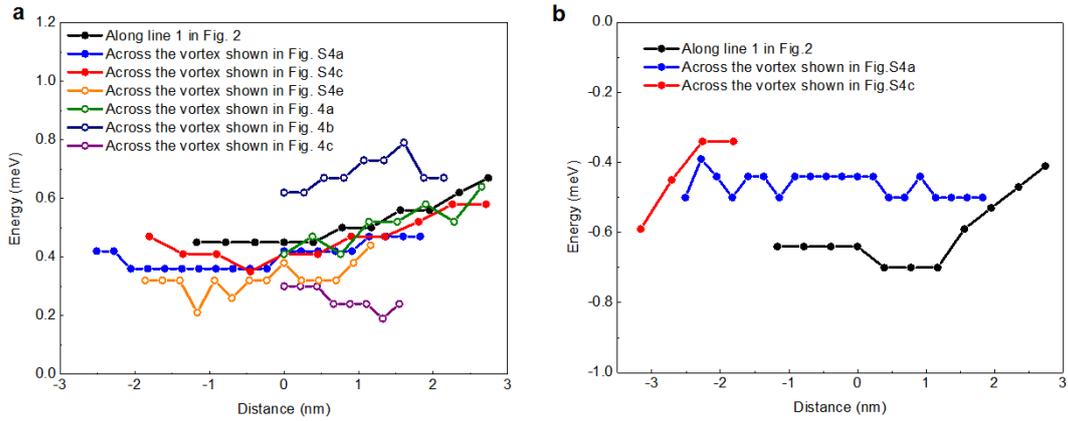

**Figure S5 | Distribution for the lowest bound state energies for different vortices.** The data represent the spatial evolution of the lowest bound state energies for positive bias (Possible $E_{+1/2}$ energy, shown in **a**) and negative bias (Possible $E_{-1/2}$ energy, shown in **b**). The empty symbols represent the peak positions for the single asymmetric bound state peaks locating at some positive energies, while the solid symbols show the data for the CdGM states existing on both positive and negative bias voltages. The absolute values of the peak position is usually larger for $E_{-1/2}$ peaks than $E_{+1/2}$ peaks.

**References**

S1. Hayashi, N., Isoshima, T., Ichioka, M. & Machida, K. Low-lying quasiparticle excitations around a vortex core in quantum limit. *Phys. Rev. Lett.* **80**, 2921-2924 (1998).

S2. Kaneko, S. *et al.* Quantum limiting behaviors of a vortex core in an anisotropic gap superconductor YNi$_2$B$_2$C. *J. Phys. Soc. Jpn.* **81**, 063701 (2012).